\documentclass[conference]{IEEEtran}
\usepackage[utf8]{inputenc}
\usepackage{amsmath}
\usepackage{amssymb}
\usepackage{microtype}
\usepackage{graphicx}
\usepackage{booktabs}
\usepackage{hyperref}
\usepackage{afterpage}
\usepackage{hyphenat}
\usepackage[font=footnotesize,labelfont=bf]{subcaption}
\usepackage{xcolor}
\usepackage{listings}
\usepackage{setspace}

\usepackage{lstlinebgrd}
\lstdefinestyle{CStyle}{
    language=C,
    breakatwhitespace=false,
    captionpos=b,
    keepspaces=true,
    numbersep=5pt,
    showspaces=false,
    showtabs=false,
    tabsize=2,
    breaklines=true,
    xleftmargin=\parindent,
    showstringspaces=false,
    basicstyle=\scriptsize\color{black},
    keywordstyle=\bfseries\color{green!40!black},
    commentstyle=\itshape\color{purple!40!black},
    identifierstyle=\color{blue},
    stringstyle=\color{orange},
    morekeywords={uint128_t,uint64_t}
}

\DeclareMathSizes{10pt}{8pt}{6pt}{6pt}

\title{Scalar Arithmetic Multiple Data:\\ Customizable Precision for Deep
Neural Networks}

\author{
  \IEEEauthorblockN{Andrew Anderson\IEEEauthorrefmark{1}, Michael J. Doyle\IEEEauthorrefmark{1}, David Gregg\IEEEauthorrefmark{1}}
  \IEEEauthorblockA{\IEEEauthorrefmark{1}School of Computer Science and Statistics\\Trinity College Dublin
  \\\{aanderso, mjdoyle, dgregg\}@tcd.ie}
}

\begin{document}

\maketitle

\frenchspacing

\begin{abstract}

Quantization of weights and activations in Deep Neural Networks (DNNs) is a
powerful technique for network compression, and has enjoyed significant
attention and success. However, much of the inference-time benefit of
quantization is accessible only through customized hardware
accelerators or with an FPGA implementation of quantized arithmetic.

Building on prior work, we show how to construct very fast implementations of
arbitrary bit-precise signed and unsigned integer operations using a software
technique which logically \emph{embeds} a vector architecture with custom
bit-width lanes in fixed-width scalar arithmetic. At the strongest level of
quantization, our approach yields a maximum speedup of $\thicksim6\times$ on
an x86 platform, and $\thicksim10\times$ on an ARM platform versus
quantization to native 8-bit integers.

%Research has shown that quantization can be effective down to the level of
%single bit representations, but without using customized hardware
%implementations, quantized DNN inference is currently restricted to the handful
%of data sizes supported in commodity hardware.

%Further, we show that entirely novel custom bit-precision vector operations can
%be defined that have no equivalent in conventional hardware vector
%architectures. We show how to use fixed-width scalar wide integer multipliers
%to construct a custom bit-precision vector convolution operation, supporting
%both signed and unsigned integer data.

%We evaluate our approach on a high-end Intel Haswell processor, and an embedded
%ARM processor. Our approach yields very fast implementations of bit-precise
%custom DNN operations, which often match or exceed the performance of operations
%quantized to the sizes supported in native arithmetic. At the strongest level of
%quantization, our approach yields a maximum speedup of $\thicksim6\times$ on the
%Intel platform, and $\thicksim10\times$ on the ARM platform versus quantization to
%native 8-bit integers.

\end{abstract}

\section{Motivation}
\label{sec:motivation}

Quantizing weights and activations in DNNs can reduce (1) the size of data,
which reduces the required memory footprint, and (2) memory traffic,  which
consumes execution time and energy~\cite{DBLP:conf/dac/HanD18}.
Performing inference in reduced bit precision also offers the possibility of
decreased execution time. Quantization is extraordinarily effective for DNNs,
and many networks can function correctly under very aggressive
quantization~\cite{DBLP:journals/corr/ZhuHMD16}.

Custom hardware accelerators~\cite{DBLP:conf/fpga/HanKMHLLXLYWYD17,
DBLP:conf/isca/ParasharRMPVKEK17, DBLP:conf/isca/HanLMPPHD16} and
FPGAs~\cite{Fu:2017} can exploit this reduction in data precision to reduce the
area and power required for the implementation of the corresponding arithmetic,
and also to increase throughput.
In contrast, conventional CPU/GPU microarchitectures typically provide native
support for just a few fixed levels of precision, such as 8-, 16-, 32- and
64-bit integer values and 32- and 64-bit floating point. The result is that
software is seldom able to take full advantage of relaxed precision
requirements.
For example, Ristretto~\cite{DBLP:journals/corr/Gysel16} can quantize DNN
activations to signed 6-bit integer precision, but we are unaware of any machine with
6-bit native arithmetic to implement the operations.

\subsection*{Contribution}
In this paper we show that bit-level custom precision on commodity hardware is
not only possible, but highly efficient for deep convolutional neural networks
(CNNs).
We extend prior research to \emph{embed} a SIMD vector architecture with custom
bit-precise integer vector lanes in native fixed-width scalar arithmetic. We
implement a wide range of operations over arrays of bit-precise signed and
unsigned integer types.

However, merely emulating the lane-wise operations of existing vector
architectures neglects the true potential of custom precision in software.
Instead, we argue that the greatest opportunities arise from defining \emph{entirely new} operations that
do not necessarily correspond to existing SIMD vector instructions.

These new operations are not easy to find because they depend on novel insights
into the sub-structure of existing machine instructions.
Nonetheless, we provide an example of one such operation: a custom
bit-level precise operation which computes the 1D convolution of two input
vectors, based on the wide integer multiplier found in general-purpose
processors.

Mixed signed-unsigned integer arithmetic is a particular challenge, and we
present a novel solution. While the performance benefits are attractive,
writing this kind of code by hand is difficult and error prone. We implement
the technique in a domain-specific code generator, which synthesizes efficient
C code implementations of the arithmetic operations.

We evaluate our approach using published quantization
scenarios on multiple hardware platforms, including embedded/IoT class devices.

\section{Background}
\label{sec:background}

Vector computer architectures are among the most successful parallel computers
for a wide variety of applications. In the 1970s supercomputers such as the
Cray-1 and Cray X-MP used deeply-pipelined vector floating point units for
performance on scientific applications. From the mid-1990s vector units started
to appear in general-purpose processors, such as Intel MMX (1996) and SSE
(1999) and PowerPC AltiVec (1997). Vector processors are single-instruction
multiple data (SIMD) parallel computers, where a single instruction operates on
multiple data points in parallel.

Modern vector processors have vector registers that contain a fixed
number of \textit{lanes}, each of which contains a scalar value.
Vector instructions operate on all lanes in parallel, using either
pipelined arithmetic units (a \textit{vector pipeline} architecture)
or using multiple parallel arithmetic units (a \textit{vector SIMD}
architecture \cite{Patterson:1990}). It is worth noting that both
\textit{vector SIMD} and \textit{vector pipeline} architectures fit
within the SIMD category of Flynn's taxonomy \cite{Flynn:1972}.

In 1997 Fisher and Dietz proposed a slightly different classification
of vector architectures \cite{Fisher:1998}. They coined the term
\textit{SIMD within a register} (SWAR) to encompass both the
then-emerging \textit{vector SIMD} architectures \textit{and} another
approach to vector parallelism that is less well known. This latter
approach is a software-only emulation of vector computing that is
implemented with scalar instructions. Scalar registers are divided
into notional sub-words, and scalar instructions are used
to operate on multiple sub-words in parallel.

Fisher and Dietz went on to create a \textit{SWAR programming model}
and compiler that could target either conventional vector instructions
or a software emulation of vectors using scalar instructions
\cite{Fisher:2003}. Unfortunately, Fisher and Dietz overloaded the
term SWAR to include three quite separate ideas: (1) all hardware
vector SIMD architectures, (2) their approach to software emulation of
vector architectures, and (3) their high-level programming model and
compiler. As a result, despite doing much pioneering work in the
field, they left no separate term for software emulation of vector
architectures. In the absence of an existing term, we refer to this
software vector approach as \textit{\textbf{S}calar \textbf{A}rithmetic with
\textbf{M}ultiple \textbf{D}ata} (SAMD).

Algorithms to operate on multiple subwords in parallel have been known
for some time. A 1957 programming textbook for the 35-bit EDSAC
contains a recursive divide and conquer algorithm to compute the
number of set bits (or \textit{popcount}), where the bit-width of each
partial sum doubles on each iteration
\cite{Wilkes:1957, Edel:2009}. The famous 1972 HAKMEM technical report
\cite{Beeler:1972} from MIT contains an example (no. 115) that
operates on base-4 subwords using bitwise operations.

In 1975 Lamport outlined a vision for processing multiple $n$-bit
``bytes'' within full word instructions (where $n$ is not necessarily
eight). Lamport outlines an algorithm for SAMD addition with and
without spacer bits, and mentions the possibility of vector times
scalar multiplication albeit without details. The main focus of the
work is lane-wise comparisons and masks, particularly for processing
character data.

% Fredriksson and Grabowski \cite{Fredriksson:2009,FREDRIKSSON:2013} use
% fast FFT convolutions for approximate string matching. To increase
% parallelism, they pack multiple symbols into a machine word, before
% computing the FFT and Fourier domain convolution on the packed words,
% which they describe as \textit{word-level parallelism}. In contrast
% our work deals with both signed and unsigned values, and computes the
% convolution in the time rather than Fourier domain.

Fu et al. present a method for multiplying two 8-bit
inputs by an 8-bit scalar value in a single step using the DSP
accelerator units on Xilinx FPGAs. According to their description ---
which is tightly bound to implementation on Xilinx FPGAs --- their
approach can deal with both signed and unsigned numbers\cite{Fu:2017} .
%
%Their method is similar to our vector scale algorithm, but it is missing the
%crucial post-pass adjustment that allows correct handling of negative inputs.
%It is not clear how their algorithm works without this step.
% It is difficult to know whether they rely on some hardware mechanism of Xilinx
% DSP units to find the correct result for negative inputs, or they use a
% post-pass adjustment similar to ours and neglect to mention it.

Umuroglu et al.~\cite{Umuroglu:2017:FFF:3020078.3021744} propose FINN, another
hardware framework for fast \emph{binarized} neural network inference. In their
formulation, points in a convolution can take on the values from the set $\{-1,
0, +1\}$. This quantization scheme can be represented without loss of
accuracy in our SAMD 2 format, which permits the values $\{-2, -1, 0, +1\}$.
Umuroglu et al. propose to use a customized hardware unit, implemented on FPGA
to actually perform inference.

\begin{figure}
\centering
\includegraphics[scale=0.25]{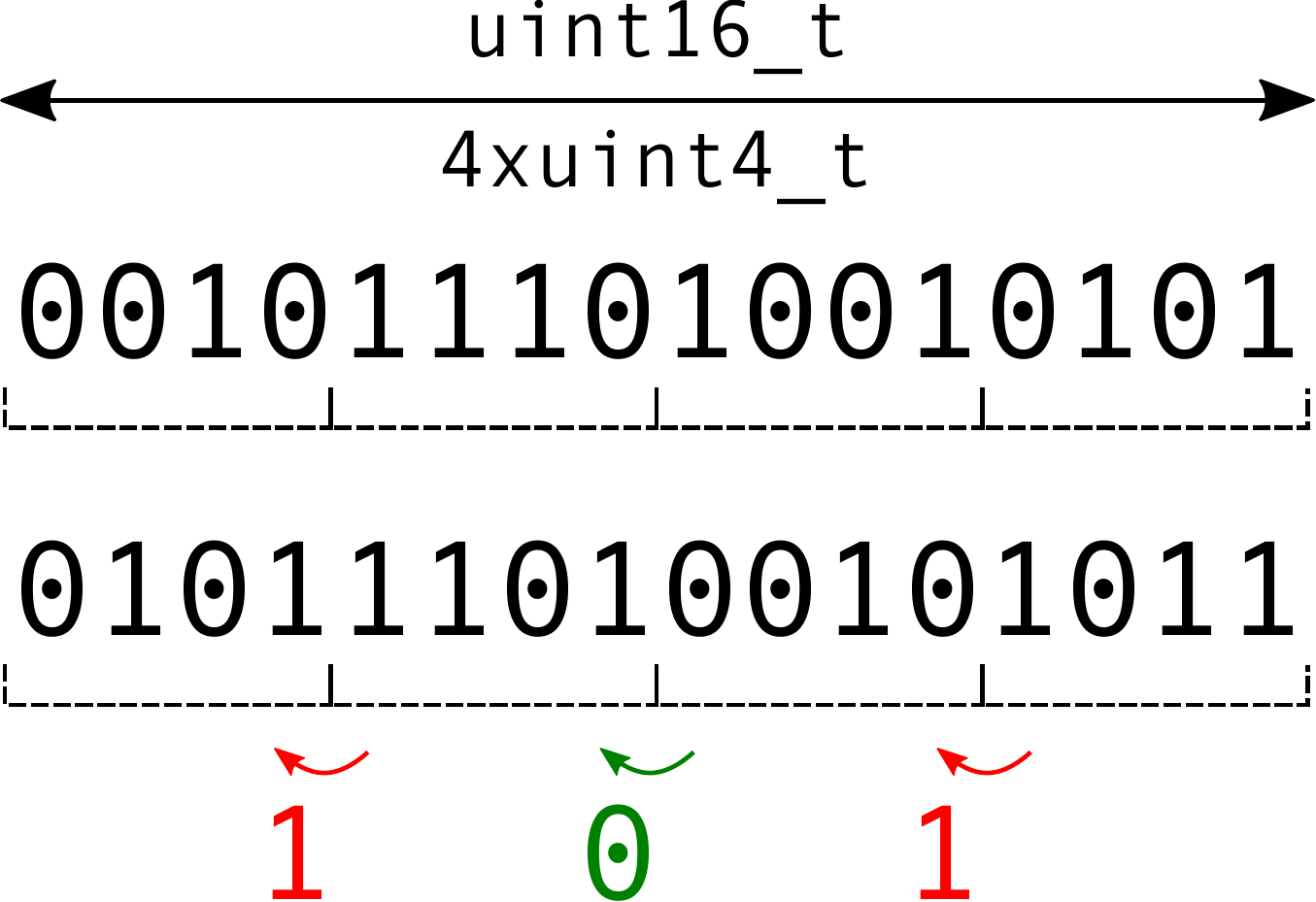}
\caption{Correct (0) and incorrect (1) carries in scalar
addition on 4-bit data when packed in a 16-bit data word. Incorrect
carries corrupt neighbouring packed values.}
\label{fig:add-scalar-simple}
\end{figure}

\subsection{Emulating Vector Architectures}

On a 64-bit 8-way vector architecture, 64-bit vector registers can be divided
into eight lanes, each containing a single byte. Vector arithmetic
instructions can be used to perform operations such as lane-wise addition.
It is equally possible to pack the same eight bytes into a 64-bit scalar
register, but applying scalar addition will not give a correct result
(Figure~\ref{fig:add-scalar-simple}). The problem is that scalar arithmetic
assumes that all bytes are parts of a singular integer value.

Scalar addition computes carries across the entire 64-bit word. In contrast,
vector operations break the carry-chain at the boundary of each lane. To
emulate vector arithmetic using scalar arithmetic, some software mechanism is
needed to sever the carries at vector lane boundaries. Fisher and Dietz
proposed \textit{spacer bits} to solve this problem.

\begin{figure}
\begin{lstlisting}[style=CStyle]
uint64_t unsigned_add_spacer(uint64_t a, uint64_t b, int bits)
{
  // create mask with zeros in spacer bits
  uint64_t mask = ~MSB_LANE_MASK(bits+1);

  // clear the spacer bits in each input
  uint64_t a_clear = a & mask;
  uint64_t b_clear = b & mask;

  // add the numbers
  return a_clear + b_clear;
}
\end{lstlisting}%\vspace{-0.2cm}
\caption{Addition of unsigned SAMD vectors using one permanent spacer
  bit between each lane}
\label{fig:unsigned-add-with-spacers}
\end{figure}

\noindent Spacer bits consist of one or more bits placed between vector lanes
to catch carries from the neighbouring lane. In the example in Figure
\ref{fig:unsigned-add-with-spacers} a single spacer bit is added between each
lane, and as a result only seven one-byte values fit within the 64-bit word
alongside the spacer bits.

To perform unsigned vector addition using unsigned
scalar addition, the spacer bits must first be set to zero.
We set the spacer bits in the inputs, $a$, $b$ to zero, using a 64-bit mask, and then add the
two resulting values. Any overflows at lane boundaries are caught in these
spacer bits. We refer to these as \textit{permanent spacer bits} because they
persist throughout the lifetime of the value.

Note that throughout this paper we use many bitwise masks to
isolate particular lanes or bits within a word. The particular
value of the mask typically depends on the lane-width and the
presence or absence of spacer bits. To describe these masks
clearly, we introduce a small function as shown in Figure
\ref{fig:build-mask}.

\begin{figure}
\begin{lstlisting}[style=CStyle]
uint64_t build_mask(int start, int len, int stride)
{
  // create a mask of len 1's
  uint64_t sub_mask = (1 << len) - 1;

  uint64_t mask = 0;
  // lay down the sub_mask at intervals of stride
   for(int i = start; i < sizeof(uint64_t)*8; i += stride){
    mask = mask | (sub_mask << i);
  }
  return mask;
}

#define MSB_LANE_MASK(w) build_mask(w-1, 1, w)
#define LSB_LANE_MASK(w) build_mask(0, 1, w)
#define ODD_LANE_MASK(w) build_mask(w, w, 2*w)
#define EVEN_LANE_MASK(w) build_mask(0, w, 2*w)
\end{lstlisting}\caption{Construct a mask for clearing bits and lanes in SAMD computation,
and examples of masks used throughout paper.}
\label{fig:build-mask}
\end{figure}

\subsection{Temporary Spacer Bits}

A downside of permanent spacer bits is that they occupy space within the vector
word. Fisher and Dietz proposed \textit{virtual spacer bits} to prevent
overflow between lanes without permanent spacer bits.

A virtual spacer bit is a short-lived spacer bit that is introduced
within a routine for operating on SAMD vectors and is eliminated before the
completion of the routine. We refer to these as \textit{temporary spacer bits}.

\begin{figure}
\centering
\includegraphics[width=0.8\linewidth]{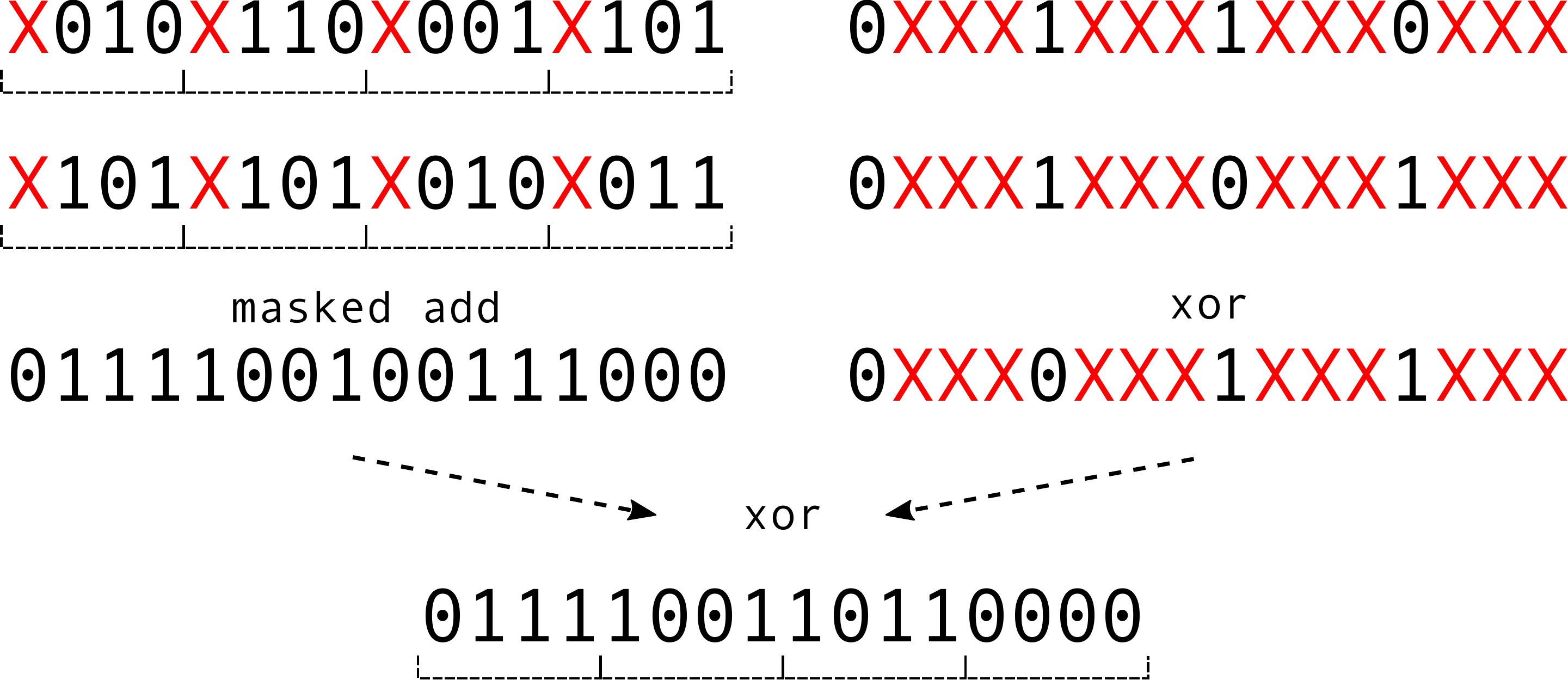}
\caption{Dataflow in SAMD addition with temporary spacer bits}
\label{fig:add-scalar-samd}
\end{figure}

Figure~\ref{fig:add-scalar-samd} shows unsigned addition using
this approach, computing the correct answer for the addition from
Figure~\ref{fig:add-scalar-simple}. The most significant bit of each lane is
masked to zero and acts as a temporary spacer bit. The scalar addition is
performed on 3-bit values, and any overflow is caught in the temporary spacer
bits.

To get a full 4-bit result in each lane, it is necessary to replace the
temporary spacer bit with the correct value of the most significant bit of the
addition. Fortunately, one-bit addition can be computed with bitwise $xor$ of
the most significant bit of each of the two input and of the carry bit that is
stored in the temporary spacer bit.

Temporary spacer bits remove the need for persistent spacer bits at
the cost of some additional computation. This greatly simplifies the task of
emulating operations found in hardware vector instruction sets such as Intel
MMX or SSE.

% Thus, we can define a common set of vector operations that might be implemented in hardware on some machines or emulated in software on others, and guarantee that the same functionality is available on all machines.

% This idea of a common set of vector operations across all target machines was an important part of the wider SWAR vision of simplifying vector programming.

%\begin{figure}
%\begin{minipage}{1\linewidth}
%\begin{small}
%\begin{verbatim}
%uint64_t unsigned_add(uint64_t a, uint64_t b, int bits)
%{
  %// create mask containing 1 in MSB of each lane
  %uint64_t mask = MSB_LANE_MASK(bits);
  %// compute the bitwise sum of MSBs of each lane
  %uint64_t msb = (a ^ b) & mask;
  %// do addition on bits-1 bits per lane
  %uint64_t sum = (a & ~mask) + (b & ~mask);
  %// add MSB sum without overflow into next lane
  %return msb ^ sum;
%}
%\end{verbatim}
%\end{small}
%\end{minipage}
%\caption{C code that performs addition of signed or unsigned SAMD vectors using temporary spacer bits}
%\label{fig:unsigned-add}
%\end{figure}

\subsection{Our Thesis}
In this paper we argue that emulating existing vector instruction sets
is exactly the wrong use of software techniques to perform vector
arithmetic with scalar instructions. Vector instruction sets are now
ubiquitous, and it is difficult for SAMD techniques to compete when
performing the same operations.

On the contrary, we argue that SAMD is most effective when it is used
for operations that are \emph{not already supported by vector
instructions}. In particular we argue that SAMD is ideal for customized
integer data precision in approximate computing
applications.
Further, as we show in Section \ref{sec:conv}, SAMD
can sometimes exploit the structure of scalar arithmetic to perform
multiple vector operations using a single scalar instruction. To
demonstrate our point, we first review examples of SAMD
operations that emulate existing vector instructions.

% However, for SAMD to be practical, we must first provide SAMD
% operations for a number of areas that Fisher and Dietz did not treat
% in their published work: signed numbers, multiplication, and a number
% of issues around the treatment of overflow.

\section{Conventional SAMD Arithmetic}
\label{sec:arith}
Fisher and Dietz published only a handful of their SAMD operations, but
the source code of Fisher's compiler for the ``SWARC''
language\footnote{http://aggregate.org/rfisher/Research/Scc/Scc.html}
contains many more. Figure \ref{fig:samd-add-sub} shows simplified
versions of SAMD operations for integer add and subtract. These
operations are correct for both signed and unsigned values, and can be
used for lane-widths with any fixed number of bits.

\begin{figure}
\begin{lstlisting}[style=CStyle]
uint64_t samd_add(uint64_t a, uint64_t b, int bits)
{
  uint64_t mask = MSB_LANE_MASK(bits);
  uint64_t msb = (a ^ b) & mask;
  uint64_t sum = (a & ~mask) + (b & ~mask);
  return msb ^ sum;
}

uint64_t samd_sub(uint64_t a, uint64_t b, int bits)
{
  uint64_t mask = MSB_LANE_MASK(bits);
  uint64_t msb = (a ^ b) & mask;
  uint64_t diff = (a | mask) - (b & ~mask);
  return msb ^ diff ^ mask;
}
\end{lstlisting}\caption{Addition and subtraction of SAMD vectors}
\label{fig:samd-add-sub}
\end{figure}

Unfortunately there is no efficient way to compute lane-wise
multiplication using scalar multiplication instructions. Unlike
addition with its single linear carry-chain which can be broken using
bitwise operations, multiplication has many carry chains. The code
generator in Fisher's SWARC compiler can generate a sequence of
bitwise shift, mask and add instructions that are capable of
performing lane-wise multiplication. Their instruction sequence is
complex and requires $O(b)$ iterations of a shift and add algorithm,
where $b$ is the lane-width. Further, each iteration uses
$O(log_2(b))$ operations to build a mask, for a total of
$O(b~log_2(b))$ operations to complete lane-wise SAMD multiplication.

In Figure \ref{fig:samd-mul} we present an improved lane-wise SAMD
multiplication sequence that requires just $O(b)$ operations for SAMD
multiplication. Our approach builds the write mask for the
multiplication in a constant number of steps. We rely on the
observation that we can efficiently compute an integer, $mask$ with
all bits in the range $start$ to $stop$ set to 1, where $start >
stop$; we simply compute $mask = (1 \ll start) - (1 \ll stop)$.

\begin{figure}
\begin{lstlisting}[style=CStyle]
uint64_t samd_mul(uint64_t a, uint64_t b, int bits)
{
  uint64_t read_mask = LSB_LANE_MASK(bits);
  uint64_t sum = 0;
  for ( int i = bits; i > 0; i-- ) {
    uint64_t bit = b & read_mask;
    uint64_t write_mask = (bit << bits) - bit;
    uint64_t to_add = a & write_mask;
    samd_add(sum, to_add, bits);
    a = a << 1;
    read_mask = read_mask << 1;
  }
}
\end{lstlisting}\caption{Improved lane-wise SAMD multiplication}
\label{fig:samd-mul}
\end{figure}

\section{Vector Scale}
\label{sec:vec-scale}
Although lane-wise vector multiplication of SAMD vectors cannot be
efficiently implemented with a scalar multiply instruction, another
useful operation can be. Key operations of many algorithms such as
matrix multiplication and discrete convolution can be expressed as
\textit{vector scale} operations, where a vector of values are
multiplied by a single scalar.

The very simplest case of vector scaling operates on unsigned lanes of
$b$ bits, where each lane is separated by $b$ permanent spacer
bits. In this case, vector scaling can be implemented by a scalar
unsigned multiplication as shown in Figure
\ref{fig:unsigned-scale-with-spacers}. In this example the scalar
value occupies the lowest $b$ bits of the $scalar$ parameter.  The
vector contains several $b$ bit values, each separated by $b$
permanent spacer bits. The product of two $b$ bit numbers contains
$2b$ bits, with the $b$ bits of overflow stored in the permanent
spacer bits.

\begin{figure}
\begin{lstlisting}[style=CStyle]
uint64_t unsigned_scale_b_spacers(uint64_t vec, uint64_t scalar, int bits)
{
  // clear the spacer bits in the input
  uint64_t mask = ODD_LANE_MASK(bits);
  uint64_t vec_clear = vec & mask;

  return vec * scalar;
}
\end{lstlisting}\caption{Unsigned vector scale with $b$ permanent spacer bits}
\label{fig:unsigned-scale-with-spacers}
\end{figure}

As we have seen for other operations, a weakness of using permanent
spacer bits is that they occupy space in the input and result.
\noindent Assuming that we want a $b$ bit result in each lane, we can simply
ignore or mask out the contents of the spacer bits.  However, it is worth
noting that the $b$ bit vector scale actually returns a product of $2b$ bits
--- the lower half in the result bits, and the upper half in the neighbouring
spacer bits.

\subsection{Vector Scale with Temporary Spacer Bits}
The downside of permanent spacer bits can be avoided by using
temporary spacer bits for vector scale as shown in
Figure \ref{fig:scale-wrap}. We first extract odd and even
numbered lanes of the vector into separate registers using bitmasks,
which creates $b$ zero-valued spacer bits between each lane. We
multiply each of the odd and even vectors by the $b$-bit scalar.  The
result is two sets of vectors where each lane contains a $2b$-bit
value. Finally, we mask off the upper half of each lane, and merge the
odd and even vectors into a single vector of $b$-bit values.

Note that when multiplying two $b$-bit integers, the result in the
lower $b$ bits is the same regardless of whether we use unsigned or
two's-complement signed multiplication \cite{Grys:2011}. Therefore,
the routine in Figure \ref{fig:scale-wrap} gives the correct $b$-bit
result in each lane for either signed or unsigned SAMD
arithmetic. Each signed lane of the vector gets the correct value,
despite the fact that the routine uses unsigned arithmetic on the
underlying integer type that is used to store the SAMD vector.

\begin{figure}
\begin{lstlisting}[style=CStyle]
uint64_t vector_scale_wrap(uint64_t vec, uint64_t scalar, const int bits)
{
  // separate odd and even numbered lanes
  uint64_t even_mask = EVEN_LANE_MASK(bits);
  uint64_t odd_mask = ODD_LANE_MASK(bits);
  uint64_t even = vec & even_mask;
  uint64_t odd = vec & odd_mask;

  // compute product and mask out higher half
  even = (even * scalar) & even_mask;
  odd = (odd * scalar) & odd_mask;
  return even | odd;
}
\end{lstlisting}\caption{Vector scale with $b$ temporary spacer bits.}
\label{fig:scale-wrap}
\end{figure}

% \subsection{Matrix multiplication}

% When multiplying matrix $A$ by a matrix $B$, we multiply all rows of A
% by all columns of $B$. This all-to-all multiplication can be performed
% using vector scale operations.

% When operating on standard data types such as 8, 16, 32 and 64-bit
% numbers, the default behaviour is that the bit-width of the result of
% an arithmetic operation is the width of the inputs. Programmers
% normally pick a sufficiently wide integer type that overflow is
% unlikely to happen. SAMD gives the programmer much more fine-grain
% control over bit widths. For example, a particular routine might
% operate on arrays of input values that the programmer knows will each
% fit within five bits. However, when pairs of these values are
% multiplied, the result is ten bits. At this point the programmer
% has a choice. They might store and operate upon the inputs as
% 10-bit values, or their might prefer a

% , allowing them to use, say, vectors of 5-bit
% numbers. Operating on arrays of 5-bit numbers can result in a
% significant saving in memory traffic as compared to 8-bit numbers.
% However, when operating on 5-bit values one is likely to be much more
% concered about overflow and the width of arithmetic results than when
% using standard types.

\section{Convolution}
\label{sec:conv}

The most computationally-intensive operation in CNNs is convolution.
Simple 1D convolution can be used to express other operations, such as
polynomial multiplication. For example, the two polynomials $2x^2 + 3x
+ 7$ and $x^2 - 5$ can be expressed as vectors $a = [2, 3, 7]$, $b =
[1, 0, -5]$. The product of the polynomials can be expressed as the
convolution $conv(a, b)$. Likewise, convolution can be expressed as
products of polynomials. Given that we already have computer hardware to perform
multiplication of numbers in a positional system, an intriguing
question arises: can we use the same hardware multipliers to perform
convolution?

The polynomial product is similar to numeric multiplication, except
that it does not allow overflow from one coefficient to the next. Therefore, if
we can \emph{prevent} overflow between coefficients packed into a scalar data
word, we can use scalar multiplication to perform the polynomial product.

\subsection{Convolution as Long Multiplication}
\label{sec:conv-mul}

A simple way to \textit{prevent} overflow between coefficients is to
ensure that there are enough bits in each lane of the result to ensure that
overflow cannot happen.

As we saw in Section \ref{sec:vec-scale}, we can scale an unsigned vector with
$b$-bit lanes by a $b$-bit unsigned scalar using native full-word
multiplication by ensuring that there are always $b$ spacer bits between each
lane of the SAMD vector. We can use a similar technique to perform convolution
of SAMD vectors, by ensuring that there are always enough bits in each result
lane to contain the maximum value computable by the convolution.

Consider the case where we convolve a vector of $b$-bit values by a 1D
convolution kernel of three $b$-bit values. Each output point in the
result consists of the sum of three values. Each of these values is
the product of a $b$-bit input value and a $b$-bit kernel value, which
gives a $2b$-bit result.

The sum of three such values requires a maximum of $2b+2$ bits to represent.
Therefore, if we guarantee that the output lane into which we compute such a
value has $2b+2$ bits, then there will never be overflow between output lanes.
One way to ensure that each output lane has $2b+2$ bits is to modify the inputs
so that each input and kernel lane has $b+1$ bits and is separated by $b+1$
spacer bits.

It is worth reiterating that our objective is not to guarantee the absence of
overflow in general, but to provide the \emph{same} overflow behaviour as a
$b$-bit scalar value. However, if $b$ bits is not enough, the programmer
needs to use a higher precision. For DNN applications, the exact precision for
quantization of each convolution is often discovered through some type of
iterative search procedure.

The effect of putting each $b$-bit value into an input lane with
$2b+2$ bits (including spacer bits) is that we can consider ourselves
to be performing arithmetic on values with base $2^{2b+2}$. By design
each input has only $b$ bits, so we are guaranteed that our
base-$2^{2b+2}$ convolution never overflows between lanes. Thus, we
can perform SAMD convolution using multiplication.

\begin{figure}
\centering
\includegraphics[width=0.48\textwidth]{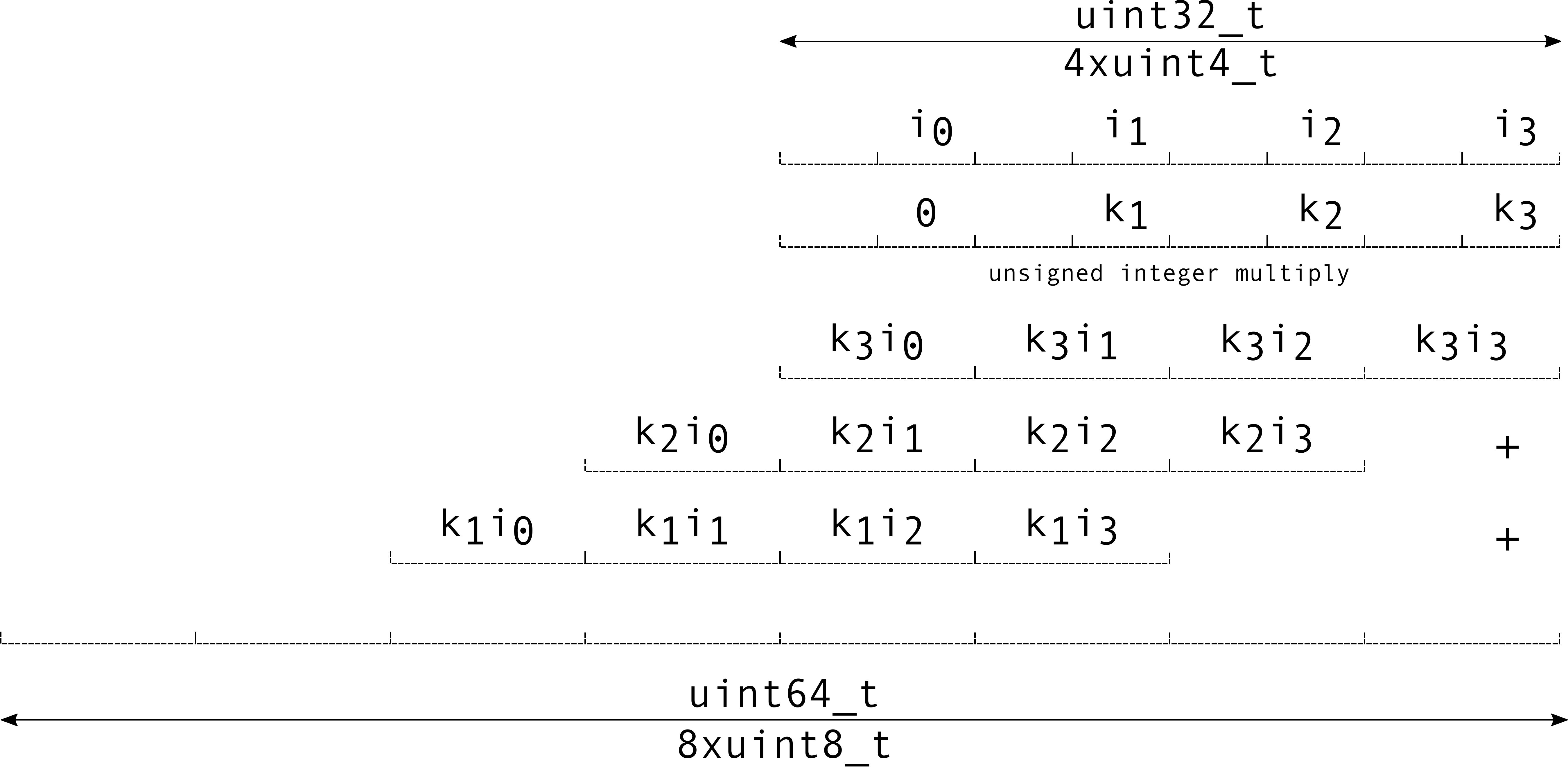}
\caption{Convolutional substructure in unsigned multiply}
\label{fig:conv-eg}
\end{figure}

Consider the example in Figure \ref{fig:conv-eg}. We have four input
values $i_0, i_1, i_2, i_3$ and a 1D convolution kernel of three values
$k_1, k_2, k_3$. The input and the kernel are packed into SAMD vectors with
zero-valued spacer lanes between each lane containing a value.  Figure
\ref{fig:conv-eg} shows each intermediate value that is computed as
part of the long multiplication.

Each intermediate value is the product of two inputs and is therefore occupies
a double-width lane in the result. When performing the multiplication we use
machine instructions that compute the full double-width result. For example,
the x86-64 architecture has a 64-bit multiply instruction that returns a
128-bit result split across two registers. The result contains the convolution
of the input and kernel.

Note that the example in Figure \ref{fig:conv-eg} has an input of
length four and kernel of length three. Each fits within a single SAMD
register, and it is therefore possible to compute the convolution with
just a single multiply. In convolutional neural networks the kernel
typically has a short, fixed length such as three or five. However,
the input is normally much longer, and we must therefore typically
compute the convolution of a kernel that fits within a single SAMD
register, and an input sequence that does not.

If we again consider the example in Figure \ref{fig:conv-eg}, but assume that
the input is just a short subsection of a much larger input, a pattern emerges.
In a long convolution with a kernel of length three, we expect each output to
consist of the sum of three products. In the example we see that two lanes of
the output --- those containing the values $k_{3}i_{0} + k_{2}i_{1} +
k_{1}i_{2}$ and $k_{3}i_{1} + k_{2}i_{2} + k_{1}i_{3}$ --- are the sum of three
non-zero products. These two output values are complete result values of the
overall convolution. In contrast, the other four outputs are partial results.
Each is the sum of just one or two products. To compute the final results for
these output points, we need to combine partial results from different
multiplications.

If we look at the pattern of non-zero intermediate values of the multiplication
in Figure \ref{fig:conv-eg} we see that they form a parallelogram-shaped
region. When we perform a longer convolution the result is a sequence of such
parallelograms. To get the result of a long convolution we simply \emph{align}
the adjacent parallelograms by bit shifting the output word and add the
overlapping points. In this way, we can build the convolution of a sequence of
arbitrary length piecewise from overlapping and adding fixed-size convolutions.

%For convolutional neural networks we must consider two additional practical
%issues. First, many CNNs use 2D convolution rather than the 1D convolution that
%we compute. However, we can compute a $3 \times 3$ 2D convolution as the sum of
%three length-3 1D convolutions. CNN convolution also typically involves
%accumulating the results of multiple 2D convolutions across \textit{channels}
%of the input.

%The overall output is computed by summing the corresponding convolution results
%across different channels. This can be computed using normal SAMD addition.
%However, rather than resolving the overlapping partial sums (or parallelograms)
%at each channel before summing the channels, it is often more efficient to
%perform the addition across channels \emph{first}, and resolve the overlapping
%partial sums later.

\section{Handling of Signed Values}
\label{sec:signed}

Our discussion of SAMD convolution and vector scale has so far assumed
that inputs are unsigned. However, CNNs normally use signed values, at
least for the convolution kernel.

Most processors provide separate signed and unsigned multiply instructions. The
lower $b$ bits of the result of multiplying two $b$-bit values is the same for
both signed and unsigned multiplication. However, the higher $b$ bits of the
result are different depending on whether the multiplication is signed or
unsigned.

If we are operating on $b$-bit inputs and computing $b$-bit outputs,
multiplication of signed values is straighforward. We simply compute
the $2b$-bit result using whatever multiplication is available and
discard the upper $b$ bits. However, multiplying $b$-bit signed
inputs to create $2b$-bit results is much more complicated, because signed
scalar multiplication deals only with the leading bit of the full machine word
as the indication of a negative value.

The issue is that the signed scalar multiplier is unable to identify signed
SAMD lanes \emph{within} a machine word. Our approach therefore uses
\emph{unsigned} scalar multiplication, with a fixup operation to adjust the
result to account for negative values in lanes. Grys \cite{Grys:2011} provides
an overview of software methods for computing signed multiplication by
adjusting the result of unsigned multiplication.

For simplicity, we assume the input vector and convolution kernels
consists of signed $b$-bit integer lanes, each separated by $b$
zero-valued spacer bits. Note that scalar times vector multiplication
is simply a special case of convolution, where the kernel contains
just one value.

\begin{figure}
\begin{lstlisting}[style=CStyle]
uint64_t samd_sign_extend4mul(uint64_t vec, int bits)
{
  // get mask with 1 in MSB of each lane
  uint64_t msb_mask = MSB_LANE_MASK(bits);
  // extract the sign bits of each lane
  uint64_t sign_mask = vec & msb_mask;
  // subtract sign bits from adjacent spacer bits
  return vec - (sign_mask << 1);
}
\end{lstlisting}\caption{Special-purpose sign extension for vector scale or convolution}
\label{fig:samd-sign-extend4mul}
\end{figure}

To sign-extend a negative lane, we need to extend its leading sign
bits into the adjacent $b$ zero-valued spacer bits to its left.
Figure \ref{fig:samd-sign-extend4mul} shows a simple method for sign
extension.
We shift the leading sign bit of each vector lane by one
position, and subtracting it from the adjacent spacer bit to the
left. Where the leading sign bit has value zero, that lane contains a
positive (or zero) number, and zero is subtracted from the adjacent
spacer bit leaving it zero.

Where the leading sign bit is one, the number is negative, and
all $b$ spacer bits must be set to one to fully sign extend the
number. We achieve this by subtracting the leading sign bit from the
adjacent spacer bit which causes it to switch from zero to one, with a
negative carry to the next bit left.
The negative carry cascades through all the spacer bits, and finally carries
into the next populated lane of the SAMD vector. As a result the next lane to
the left of a negative number is reduced by one.

On first view one might expect this change of value to
catastrophically change the result of the overall computation, but
remarkably it does not. When we multiply the resulting vector value by
the scalar, we get an answer that is very close to the correct one.
The rightmost lane of the resulting vector is always correct. Other
lanes have the correct value if the lane to the left is non-negative.
Where the lane immediately to the left is negative, the current lane
has a value one less than the true value. We can correct this by adding 1
to each of the lanes with a negative value to the left.

\begin{figure}
\begin{lstlisting}[style=CStyle]
uint128_t signed_convolution(uint64_t vec, uint64_t kernel, int bits)
{
  // compute 128 bit product of 64-bit inputs
  uint128_t product = (uint128_t)vec * (uint128_t)scalar;

  // fix underflow from adjacent lane
  uint128_t mask = MSB_MASK(bits*2);
  uint128_t sign_bits = product & mask;

  // add the sign bit to itself to increment
  // lane to the left of a negative number
  product = product + sign_bits;

  // correct the MSB for wrongly removed sign
  return product ^ sign_bits;
}
\end{lstlisting}\caption{Signed convolution algorithm with underflow correction}
\label{fig:conv-2b}
\end{figure}

Figure \ref{fig:conv-2b} shows code to perform the convolution and
correct the underflow that propagates from adjacent negative lanes.
The shown code is efficient, but non-obvious. The obvious way to
correct the underflow from an adjacent negative lane is to extract the
most significant bit (which is 1 in the case of a negative number),
shift the bit one position to the left so it aligns with least
significant bit of the lane to the left and add. However, the obvious
approach contains a subtle bug.

Where the correct value in an output
lane is zero, the negative underflow from an adjacent lane pushes the
value to -1. In two's complement the representation of -1 is all
1's. If we correct this value by adding one, the lane will get the
correct value, but will also overflow into the next lane.
This would be correct except that the next lane to the left will have
already been incremented because its direct neighbour is negative
before the increment. Therefore, the value is incremented twice.

We instead use a non-obvious sequence that extracts the sign bits, but
rather than shifting them left adds them to the same location. Where
the sign bit is 1, this causes an overflow into the adjacent lane.
Where the lane contains the value -1, the overflow switches it to
zero, but when we add another 1 to the MSB there is no overflow into
the neighbour. Finally, we correct the MSB of each output lane.

\subsection{Optimizations for Permanent Spacer Bits}

The method of signed convolution presented so far has assumed that we
need to compute the correct value for the MSB of each vector
lane. However, it is worth noting that computing the correct MSB can
be surprisingly expensive for SAMD operations.

For example, in the SAMD addition operation in Figure \ref{fig:add-scalar-samd},
half of the bitwise operations are devoted to computing the correct value of
the MSB. If we are willing to sacrifice one bit of precision and not maintain
the MSB, SAMD addition is much more efficient.
Where we maintain a permanent spacer bit in the MSB of each SAMD lane, we can
use cheaper operations, such as the addition in Figure
\ref{fig:unsigned-add-with-spacers} which uses far fewer bitwise operations.

In the case where we maintain a permanent spacer bit in the MSB of
each lane, we can also perform convolution more efficiently.  First,
the final \textit{xor} operation in the convolution code in Figure
\ref{fig:conv-2b} is unnecessary if we do not need to compute the
MSB. Given that the \textit{xor} operates on a double-width value (in
this case 128 bits), this likely corresponds to two scalar machine
instructions.

A second optimization of convolution arises because of the way that we
correct for negative adjacent lanes. Recall that to correct for
negative lanes we extract the sign bit from the result of the
multiplication and add it back into the result. Where the sign bit has
value 1, this propagates a positive carry into the neighbouring lane.
Unfortunately this operation also corrupts the sign bit, but where the
MSB is a spacer bit, this is not a problem. However, the result of the
addition can leave the lane of a SAMD vector in one of two possible
states.

If before the addition, the SAMD lane contained the value -1,
then after the addition, the SAMD lane contains the value zero in all
bits except the MSB, which has value 1. If before the addition the
SAMD lane contained any other value, then after the addition it
contains a zero in the MSB and arbitrary values in other bits.

In CNN convolution, the output of our convolution operation is normally
added to an accumulation of the values across multiple channels. Where
a permanent spacer bit is used in SAMD lanes, we normally clear the
MSB of each operand before performing the addition (as shown in Figure
\ref{fig:unsigned-add-with-spacers}).

However, given that the MSB of each lane in the output of our convolution
operation is already zero all cases, except the case where it is one and all
other bits are zero, addition with a SAMD lane where the MSB is zero cannnot
overflow into the neighbouring lane.
Therefore, where the output of our convolution feeds directly into addition
with a permanent spacer bit, there is no need to clear the MSB of the lanes of
our convolution result.

\section{Experimental Evaluation}

To implement low precision integer arithmetic on a general purpose processor
without hardware support, the simplest option available to the programmer is to
\emph{unpack} the low-precision values into the closest available native
integer type. On the vast majority of commodity processors, the smallest
integer type is 8 bits in length. For example, the pointwise sum of two
streams of 6-bit integers can be computed by zero-padding the values to 8 bits,
and using native 8-bit addition.

Using this approach, the arithmetic sequence performing any computation on
streams of values smaller than 8 bits is identical, with the only adjustment
being the number of zeroes used to inflate individual elements (i.e. the shift
factor, on machines which shift in zeroes).

The class of convolutions with inputs of width at most 8 bits is precisely the
target of our evaluation, since these are typically only used when hardware
acceleration is available. Figure~\ref{fig:dnn-convolution} shows some example
C code implementing DNN convolution. For our evaluation, we built a small
code generator which synthesizes the SAMD instruction sequence implementing the
hilighted arithmetic, using the underflow correction approach from
Figure~\ref{fig:conv-2b}. Since unpacking to native 8-bit integer arithmetic
can be used for all narrower bitwidths, the performance of the native 8-bit
version of the code is our baseline for benchmarking.

After we generate the C code implementing the SAMD arithmetic operations, we
compile everything to machine code using the native compiler (in our case, GCC
release 7.2).

\begin{figure}[h]
\begin{lstlisting}[style=CStyle,linebackgroundsep=-0.06\linewidth,linebackgroundwidth=0.72\linewidth,linebackgroundcolor={\ifnum\value{lstnumber}>5\ifnum\value{lstnumber}<7\color{blue!30}\else\ifnum\value{lstnumber}<12\color{orange!30}\fi\fi\fi}]
for (unsigned m = 0; m < kernels; m++)
 for (unsigned h = k/2; h < (img_h-(k/2)); h++)
  for (unsigned w = k/2; w < (img_w-(k/2)); w++) {
   T result = output[m][h][w];
   for (unsigned d = 0; d < channels; d++) {
    for (unsigned y = 0; y < k; y++)
     for (unsigned x = 0; x < k; x++)
      result +=
      (T)image[d][(h + y)-(k/2)][(w + x)-(k/2)]
      *
      (T)kernel[m][d][y][x];
   }
   output[m][h][w] = result;
 }
\end{lstlisting}

\caption{\textbf{DNN convolution} (\textsc{direct-sum2d}). \\The entire coloured
area corresponds to a single $k \times k$ 2-dimensional convolution, and the
orange area specifically corresponds to the 1-dimensional sub-convolution which
is implemented by the SAMD convolution operator.}

\label{fig:dnn-convolution}
\end{figure}

\subsection{Experimental Setup}

We evaluated our approach on an AMD Ryzen 7 2700X, and on an ARM Cortex-A57.
For our experiments we are using the convolutional layers of the VGG-B network
of Simonyan et al.~\cite{DBLP:journals/corr/SimonyanZ14a}. From left to right
on the graphs, the depth of the layer in the network increases, increasing the
number of channels and the number of convolutional kernels. The precise
dimensions of each convolution are tabulated
in~\cite{DBLP:journals/corr/SimonyanZ14a}.

%On our figures, the native direct convolution algorithm \textsc{direct-sum2d}
%shows the performance of the convolution layer when quantized to native 8-bit
%signed integer weights and activations using the same convolution algorithm.
%
The \emph{SAMD(N)} bars show the performance of our generated SAMD convolution
layer when quantized to $N$-bit signed integer weights and activations. In
experiments, we step $N$ down from $8$ (emulating native 8-bit integers) to $2$
(a single data bit with a sign bit).

Recall from Figure~\ref{fig:conv-eg} that each individual multiply instruction
generated to implement the SAMD arithmetic sequence computes \emph{multiple}
subterms of the overall sum-of-products in a $k \times k$ 2-D convolution. If
$I$ inputs are convolved with a $K$ point kernel, the SAMD arithmetic uses only
$(I - K) + 1$ integer multiplies and $(I - K)$ integer additions to compute the
result, while the native arithmetic must use $(I \times K)$ multiplies and adds
to compute the result. Since the length of the machine word is fixed, $I$ is
larger for smaller types. Using 3-bit integers with 64-bit machine words, a
3-point 1D convolution can be computed using only $(10 - 3) + 1 = 8$
multiplies using SAMD arithmetic, while the native 8-bit code requires $30$
multiplies.

\subsection{Experimental Results: Ryzen 7}

\begin{figure}
  \centering
  \includegraphics[width=0.9\linewidth]{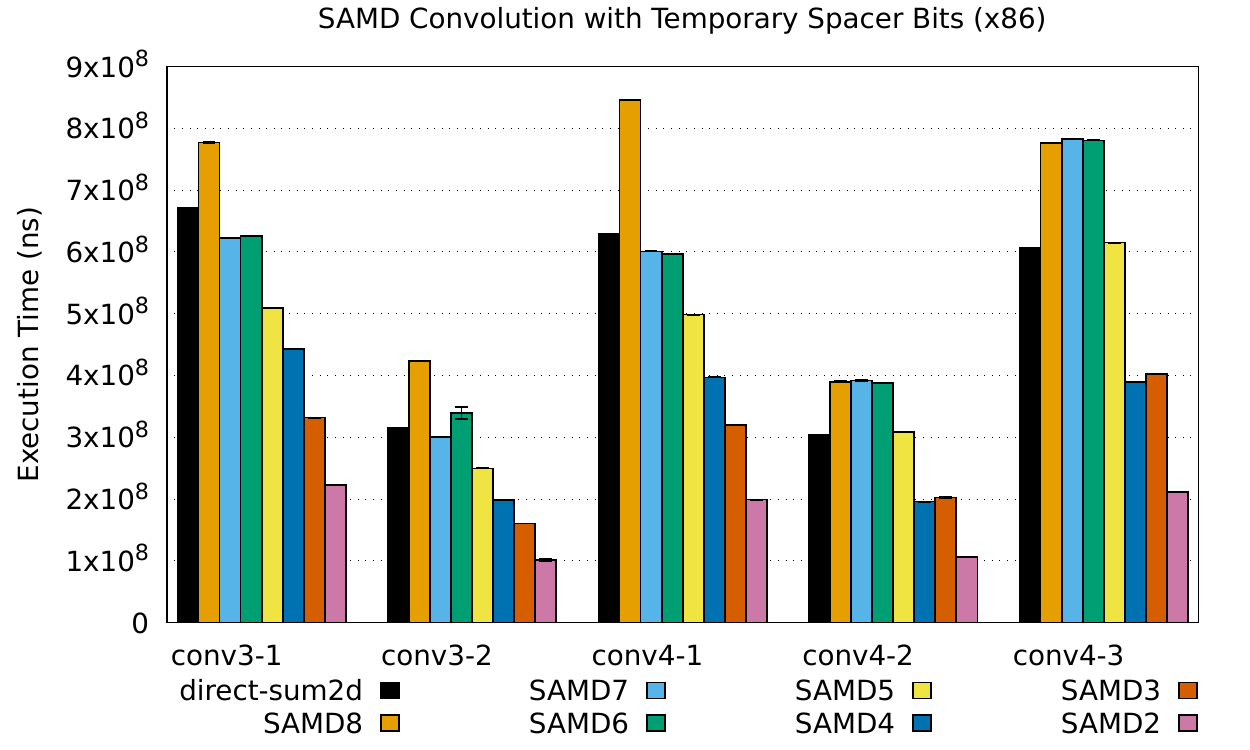}
  \caption{\textbf{AMD Ryzen 7 2700X}: Temporary Spacer Bits}
  \label{fig:results-x86-temp}
\end{figure}

\begin{figure}
  \centering
  \includegraphics[width=0.9\linewidth]{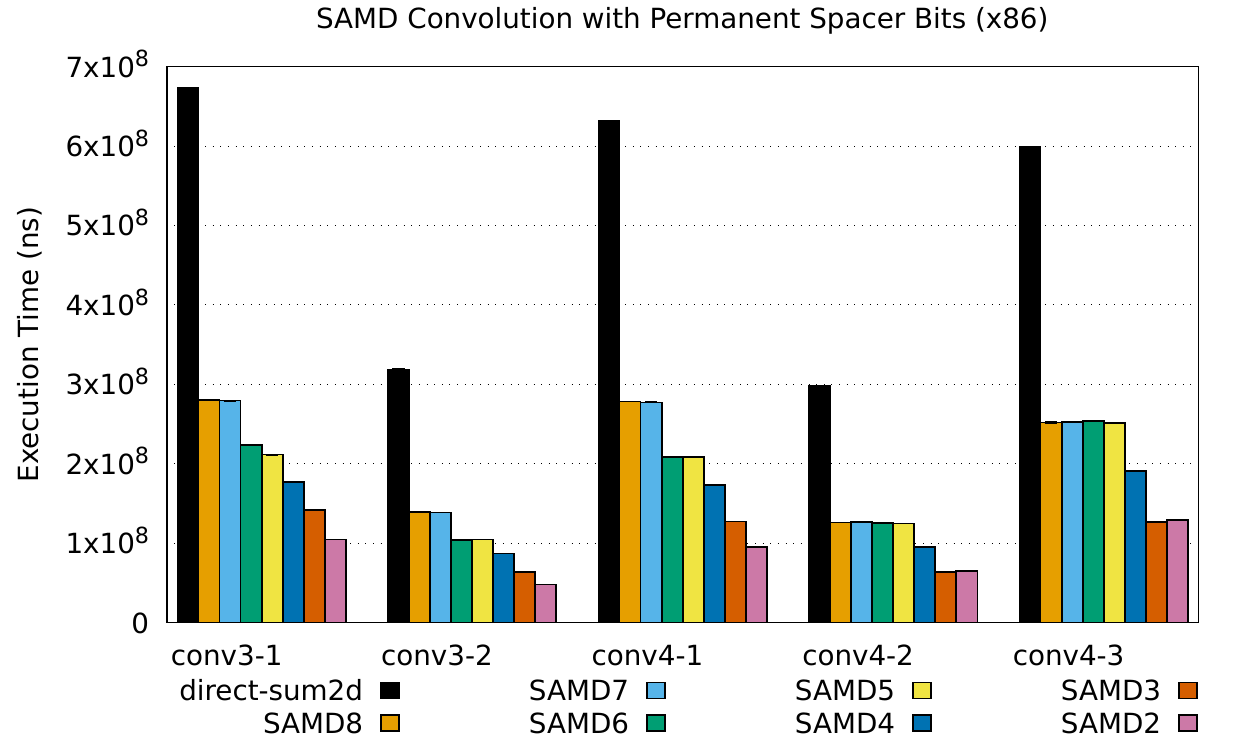}
  \caption{\textbf{AMD Ryzen 7 2700X}: Permanent Spacer Bits}
  \label{fig:results-x86-perm}
\end{figure}

Figures~\ref{fig:results-x86-temp} and~\ref{fig:results-x86-perm} show the
results of benchmarking on Ryzen 7.
Using temporary spacer bits, a more complex instruction sequence is required
(shown in Figure~\ref{fig:add-scalar-samd}) than where the format contains
permanent spacer bits, leading to a higher execution time overall.
Although using temporary spacer bits minimizes memory, the
overhead of the more complex instruction sequence is clearly visible in
the execution time.

The Ryzen 7 processor has fast native SIMD vector instructions for 8-bit
integer arithmetic. In convolution layers late in the VGG network, with very
large numbers of channels, the native 8-bit SIMD code is faster than some of
the wider SAMD instances with temporary spacer bits, but in general, as the
quantization factor increases (meaning we can fit more values in a SAMD vector)
the performance of SAMD improves.

The best available speedup for an individual convolution
layer, using 2-bit quantization, is approximately $6\times$.

\subsection{Experimental Results: ARM Cortex-A57}

\begin{figure}
    \centering
    \includegraphics[width=0.9\linewidth]{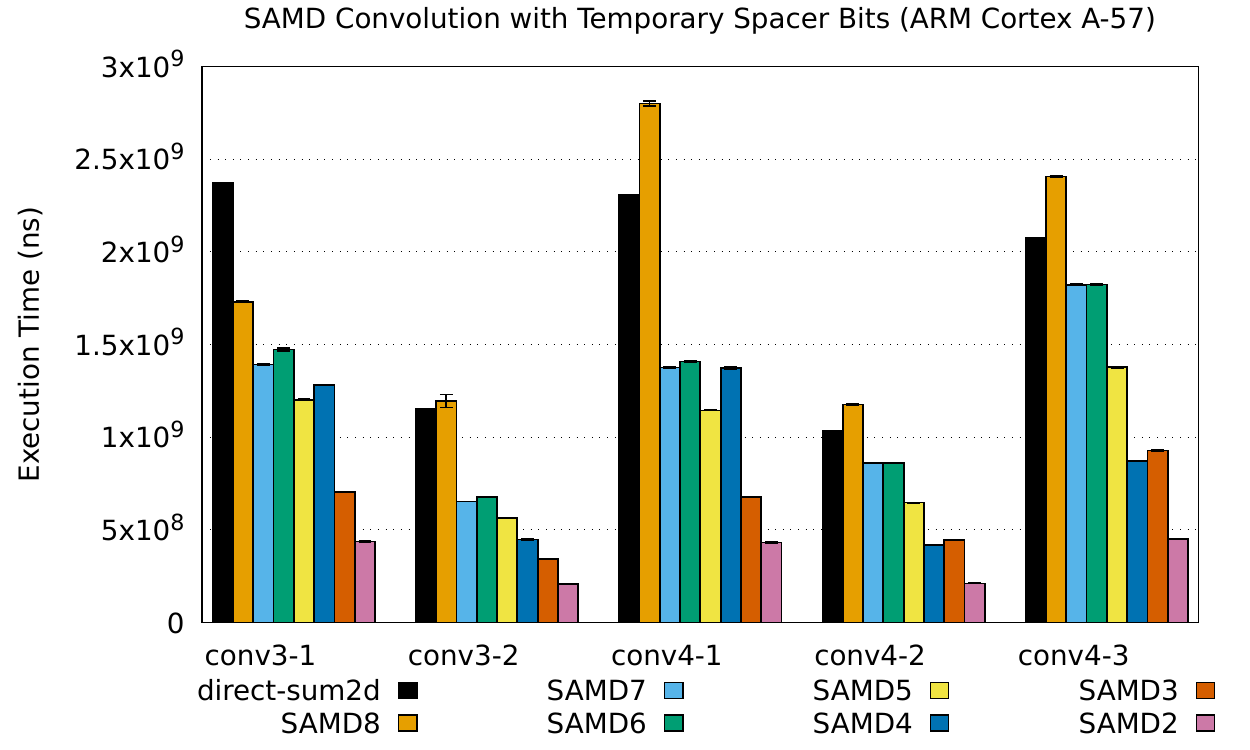}
    \caption{\textbf{ARM Cortex-A57} Temporary Spacer Bits}
    \label{fig:results-a57-temp}
\end{figure}

\begin{figure}
  \centering
  \includegraphics[width=0.9\linewidth]{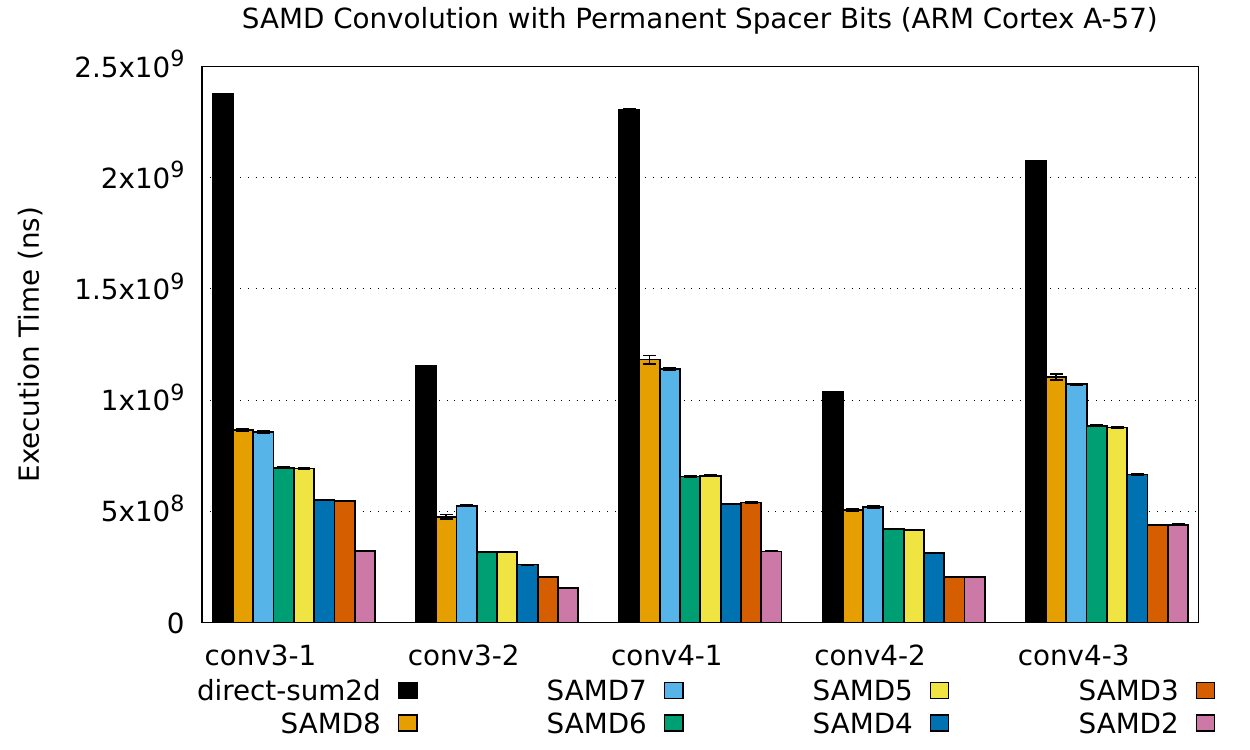}
  \caption{\textbf{ARM Cortex-A57} Permanent Spacer Bits}
  \label{fig:results-a57-perm}
\end{figure}

On the ARM Cortex-A57, the SAMD code for temporary spacer bits is faster than
unpacking to native integers for all SAMD widths less than 8 bits (Figure~\ref{fig:results-a57-temp}).
This is likely due to the narrower issue width and increased operation latency
versus the Haswell processor.
Using permanent spacer bits (Figure~\ref{fig:results-a57-perm}), the gap widens
even further. Where the incoming data are in reduced precision, SAMD is a very
effective mechanism to achieve parallelized execution on constrained processors
such as the Cortex-A57.

Using the low-complexity operations with permanent spacer bits
(Figure~\ref{fig:results-a57-perm}), the best speedup available for an
individual SAMD convolution layer, using 2-bit quantization, is approximately
$10\times$ versus native 8-bit integer implementation.

\section{Conclusion}

Our approach allows proven quantization schemes from prior work to be run on a
general-purpose CPU in a highly efficient manner. As we have demonstrated, the layers
of this network can be run on a Cortex-A57 with a 2-bit quantized inference
scheme yielding a speedup approaching $10\times$ for several layers.

As demonstrated by our experimental evaluation, quantization can enable not
only memory savings, but also inference performance gains without
requiring custom hardware support. Software support for bit-precise quantized
DNN operations using our approach can often match, and sometimes far exceed,
the performance of inference quantized to the sizes supported in native
arithmetic.

\subsection*{Acknowledgment}

\begin{scriptsize}
\noindent This project has received funding from the European Union's
Horizon 2020 research and innovation programme under grant agreement No 732204
(Bonseyes). This work is supported by the Swiss State Secretariat for
Education, Research and Innovation (SERI) under contract number 16.0159.
This work was supported by Science Foundation Ireland
grant 12/IA/1381. This work was supported in part by
Science Foundation Ireland grant 13/RC/2094 to Lero --- the Irish Software
Research Centre (www.lero.ie).
The opinions expressed and arguments employed herein do not necessarily reflect the
official views of these funding bodies.
\end{scriptsize}

\vfill

\bibliographystyle{plain}

\bibliography{references}

\end{document}